\documentclass[aps,prb,superscriptaddress,showpacs, 
twocolumn
]{revtex4}
\usepackage{graphicx}
\usepackage{epstopdf}
\usepackage{amssymb,amsmath,amsfonts}

\graphicspath{ {images/} }
\DeclareGraphicsExtensions{.png,.pdf}

\begin{document}
\title{Doping-induced quantum cross-over in Er$_2$Ti$_{2-x}$Sn$_x$O$_7$}
\author{M. Shirai}
\affiliation{London Centre for Nanotechnology and Department of Physics and Astronomy, University College London, 17-19 Gordon Street, London WC1H 0AJ, UK}
\author{R.S. Freitas}
\affiliation{Instituto de Fisica, Universidade de Sao Paulo, CP 66318, 05314-970 Sao Paulo, SP, Brazil}
\author{J. Lago}
\affiliation{Department of Inorganic Chemistry, Universidad del Pais Vasco (UPV-EHU), 48080 Bilbao, Spain}
\author{S.T. Bramwell}
\affiliation{London Centre for Nanotechnology and Department of Physics and Astronomy, University College London, 17-19 Gordon Street, London WC1H 0AJ, UK}
\author{C. Ritter}
\affiliation{Institut Laue Langevin, Boite Postale 156X, F-38042 Grenoble 9, France}
\author{I. \v Zivkovi\'c}
\email{ivica.zivkovic@epfl.ch}
\affiliation{Laboratory for Quantum Magnetism, Institute of Physics, EPFL, CH-1015 Lausanne, Switzerland}

\date{\today}

\begin{abstract}
We present the results of the investigation of magnetic properties of the Er$_2$Ti$_{2-x}$Sn$_x$O$_7$ series. For small doping values the ordering temperature decreases linearly with $x$ while the moment configuration remains the same as in the $x = 0$ parent compound. Around $x = 1.7$ doping level we observe a change in the behavior, where the ordering temperature starts to increase and new magnetic Bragg peaks appear. For the first time we present evidence of a long-range order (LRO) in Er$_2$Sn$_2$O$_7$ ($x = 2.0$) below $T_N = 130$ mK. It is revealed that the moment configuration corresponds to a Palmer-Chalker type with a value of the magnetic moment significantly renormalized compared to $x = 0$. We discuss our results in the framework of a possible quantum phase transition occurring close to $x = 1.7$.
\end{abstract}


\maketitle

%
%
%
%


Compounds based on the pyrochlore lattice of rare-earth ions, with a general formula \textit{A}$_2$\textit{B}$_2$O$_7$ (\textit{A} = rare-earth, \textit{B} = metal), have in recent decades provided a fruitful arena for discoveries of new emergent phenomena based on the interplay between exchange interactions, single-ion anisotropies, geometrical frustration and quantum fluctuations. The most notable example represents the case of a spin-ice manifold where excitations in the form of magnetic monopoles have been established~\cite{Fennell2009}. A necessary prerequisite for the spin-ice ground state is the Ising-type of a single-ion anisotropy found in Ho$_2$Ti$_2$O$_7$~\cite{Fennell2009}, Dy$_2$Ti$_2$O$_7$~\cite{Ramirez1999}, as well as in an erbium-based spinel compound CdEr$_2$Se$_4$~\cite{Lago2010}. To the contrary, the Er-based titanate, Er$_2$Ti$_2$O$_7$, exhibits an almost ideal \textit{XY}-anisotropy, with moments confined to the plane perpendicular to local $<$111$>$ axes. 

Er$_2$Ti$_2$O$_7$ has the highest ordering temperature of all the rare-earth pyrochlores studied so far ($T_N = 1.23$ K, Ref.~\cite{Reotier2012}) with no apparent sample dependence, indicating a robust ground state. It orders into a non-coplanar $k = 0$ antiferromagnetic structure, described by the $\psi_2$ basis vector of the $\Gamma_5$ irreducible representation~\cite{Poole2007}. Interestingly, another spin configuration ($\psi_3$) within $\Gamma_5$ results with the same energy as $\psi_2$ ($\psi_1$ and $\psi_2$ in Ref.~\cite{Guitteny2013}). It has been long-proposed that $\psi_2$ is favored based on entropic grounds with both thermal~\cite{Villain1980} and quantum fluctuations~\cite{Champion2003}. This so-called 'order-by-disorder' scenario has recently been verified both experimentally~\cite{Savary2012} and theoretically~\cite{Zhitomirsky2012}. It is thus of great interest to investigate further the delicate balance between these degenerate states.

One direction of investigation is a magnetic dilution of erbium sites by non-magnetic ytrium ions, Er$_{2-x}$Y$_x$Ti$_2$O$_7$. Niven \textit{et al}.~\cite{Niven2014} have concluded from a heat capacity study that the dilution reduces the ordering temperature in a linear fashion, with a percolation threshold of $\approx 60$ \%. To the contrary, two independent theoretical studies~\cite{Maryasin2014,Adreanov2015} suggested an existence of a first order phase transition between $\psi_2$ and $\psi_3$ at around 7 \% level. Indeed, recent neutron diffraction experiments~\cite{Gaudet2016} showed an instability of $\psi_2$ state and closing of an energy gap at somewhat higher level of doping, between 10 \% and 20 \%. They interpreted the results on the 20 \% sample as a frozen mosaic of $\psi_2$ and $\psi_3$ domains.

The emergence of the $\psi_3$ state is explained through the 'order-by-structural disorder' mechanism which competes with the before-mentioned, entropy-based order-by-disorder~\cite{Maryasin2014}. When the system is found within such an intricate balance between the states, it is natural to consider effects of exchange energy tuning. A very general theoretical approach has been conducted by Wong \textit{et al}.~taking into account the anisotropic nearest-neighbor couplings~\cite{Wong2013}. Due to their very similar ionic radius, magnetic dilution with ytrium does not affect much the exchange interaction between erbium moments~\cite{Niven2014,Gaudet2016}. On the other hand, when doping is performed on the \textit{B}-site of the pyrochlore lattice, replacing Ti by either Sn or Ge, the effects are much more pronounced. For instance, it has been claimed recently that in Er$_2$Ge$_2$O$_7$ the order-by-disorder mechanism favors the $\psi_3$ state, albeit the study has been performed on powder samples, with the final confirmation from single crystals still awaited~\cite{Dun2015}.

Relatively more work has been done on various rare-earth pyrochlore stannates. In Ho$_2$Sn$_2$O$_7$ and Dy$_2$Sn$_2$O$_7$ the low-temperature features characteristic of a spin-ice behavior have been reported, similar to their Ti-based counterparts~\cite{Matsuhira2004}. With \textit{A} = (Tb, Yb) the situation is quite different. While there are no signatures of magnetic ordering down to lowest temperatures in Tb$_2$Ti$_2$O$_7$~\cite{Gardner1999}, making it a spin-liquid candidate, a clear LRO has been detected in Tb$_2$Sn$_2$O$_7$~\cite{Mirebeau2005}, with persistent magnetic fluctuations exhibiting dynamical freezing within the ordered state~\cite{Dahlberg2001}. For Yb compounds an additional complication arises from a strong sample dependence where a small Yb-deficiency~\cite{Chang2012} or 'stuffing' of Yb on the Ti-site~\cite{Ross2012} in Yb$_2$Ti$_2$O$_7$ quickly suppresses a first-order transition around 240 mK. Similarly for Yb$_2$Sn$_2$O$_7$ two recent reports have shown a transition to a  highly dynamic splayed ferromagnetic structure below 140 mK~\cite{Yaouanc2013,Lago2014}. It has been argued that the system is very susceptible to even the smallest amounts of disorder due to the proximity of a quantum critical point, resulting in a cluster glass-like state~\cite{Lago2014}.

Er$_2$Sn$_2$O$_7$ has been a subject of several studies, all of them reporting an absence of a LRO~\cite{Lago2005,Sarte2011,Guitteny2013}. In muon-spin relaxation experiments~\cite{Lago2005} the relaxation rate showed the same qualitative slowing down of spin fluctuations as observed in the Ti analog. On the other hand, the measured spectra showed the same exponential type of relaxation at high and low temperatures, indicating no ordering. Low temperature heat capacity data (down to 350 mK) observed a sharp rise above the level expected from the assumed crystal field scheme~\cite{Sarte2011}. It has been hypothesized that the cause of the increase might come from either the split of the lowest lying crystal field doublet or from the phase transition occurring just below 350 mK. However, inelastic scattering profile at 100 mK did not reveal the presence of additional magnetic Bragg peaks which would indicate LRO. Lastly, magnetization measurements~\cite{Guitteny2013} down to 100 mK showed no sign of ordering but the imaginary component of ac susceptibility did reveal evidence of freezing below 200 mK. It is noteworthy to mention that from the magnitude of the shift of its maximum one can infer the existence of superparamagnetic clusters, rather than a classical spin-glass behavior~\cite{Mydosh1993}. Indeed, neutron scattering experiments in Ref~\cite{Guitteny2013} revealed the presence of short-range correlations at 1.5 K, which they have attributed to a different irreducible representation, $\Gamma_7$, with the basis vector corresponding to a particular spin configuration called Palmer-Chalker~\cite{Palmer2000}.

The evidence of a different irreducible representation characterizing Er$_2$Sn$_2$O$_7$ indicates a possible quantum phase transition induced by doping of Er$_2$Ti$_2$O$_7$. In that context we present the results of our study on several members of the series Er$_2$Ti$_{2-x}$Sn$_x$O$_7$ using heat capacity, ac susceptibility and neutron scattering. We have found that the ordering temperature decreases linearly with $x$ down to $x = 1.7$, characterized by the same basis vector $\psi_2$ of the $\Gamma_5$ representation as the parent compound $x = 0$. For the $x = 2$ end member of the series we report for the first time evidence of a LRO below 130 mK, with a spin configuration corresponding to a Palmer-Chalker type. 

%
%
%
%


Polycrystalline samples were prepared via the standard solid-state method. Stoichiometric amounts of Er$_2$O$_3$, TiO$_2$ and SnO$_2$ where thoroughly ground and fired at 1200~$^0$C and 1400~$^0$C with intermediate regrinding. The quality and phase purity of the samples was confirmed by in-lab X-ray and neutron diffraction measurements. Neutron diffraction data were collected in the temperature range $0.05 < T < 1.25$ K at the D20 instrument at the Institute Laue Langevin (ILL, France). Inelastic scattering has been performed at the HET spectrometer at ISIS (Oxford, UK). The specific heat data were obtained with a Quantum Design Dynacool system equipped with a dilution refrigerator option using a standard semi-adiabatic heat pulse technique. The addendum heat capacity was measured separately and subtracted. AC susceptibility was obtained using home-made set of induction coils on a dilution refrigerator.



%
%
%
%

\begin{figure}[t]
	\includegraphics[width=0.45\textwidth]{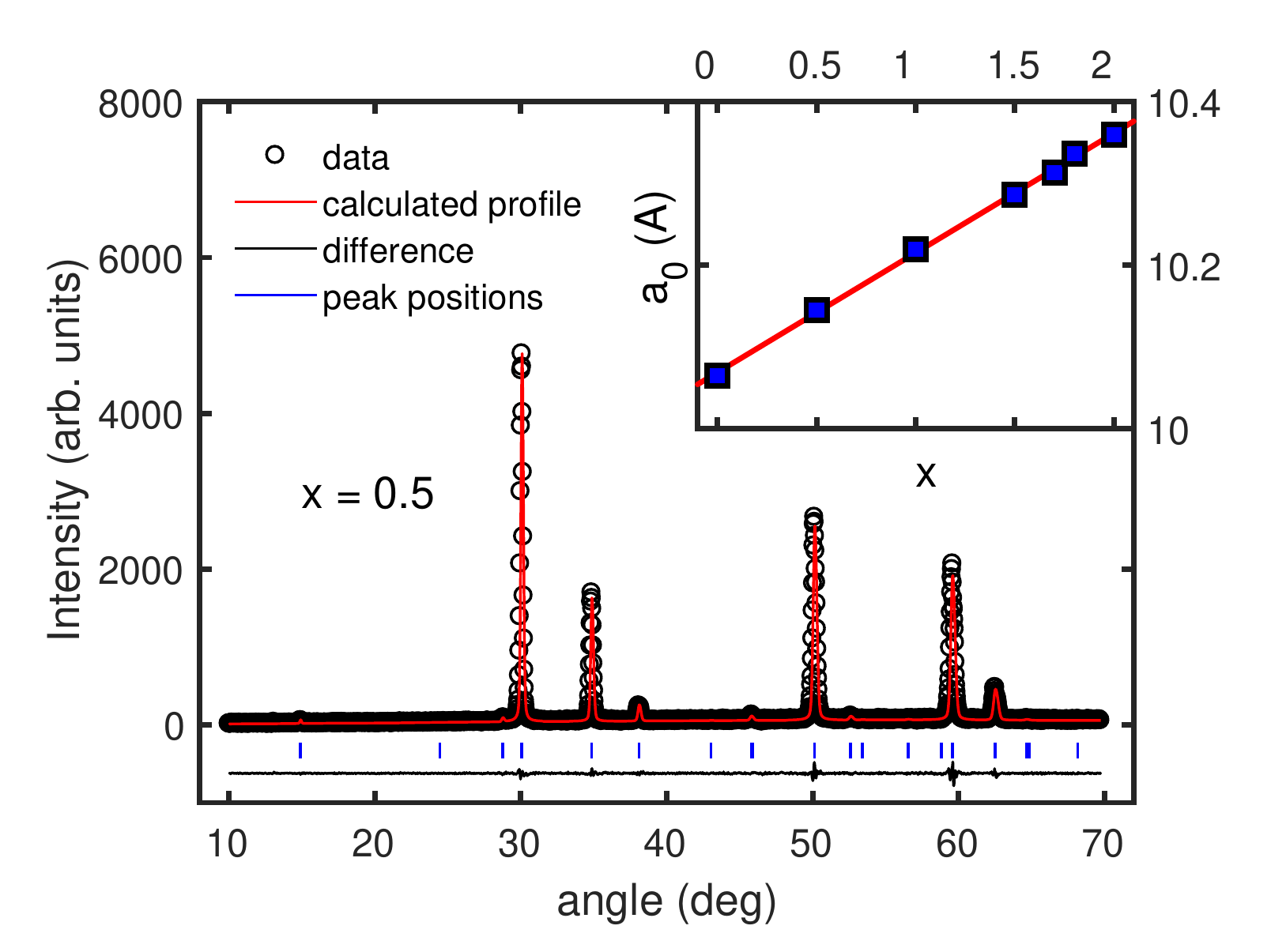}
	\caption{(Color online) x-ray diffraction pattern for the $x = 0.5$ composition. Peak positions are marked with small vertical bars. The line at the bottom indicates the difference between the measurement and the model prediction. Inset shows the doping dependence of the unit cell parameter $a_0$ across the series. The straight line indicates that investigated compounds follow Vegard´s law~\cite{Vegard1921}.}
	\label{fig-diffraction}
\end{figure}

In Fig.~\ref{fig-diffraction} we present the powder diffraction pattern for $x = 0.5$ as an example. All the peaks could be indexed according to the calculated profile for the $Fd\overline{3}m$ space group, and no additional phases could be detected. The unit cell size, given by the parameter $a_0$, increases linearly with $x$ due to a larger ionic radius of Sn compared to Ti, see inset of Fig.~\ref{fig-diffraction}.

%
%

\begin{figure}
	\includegraphics[width=0.45\textwidth]{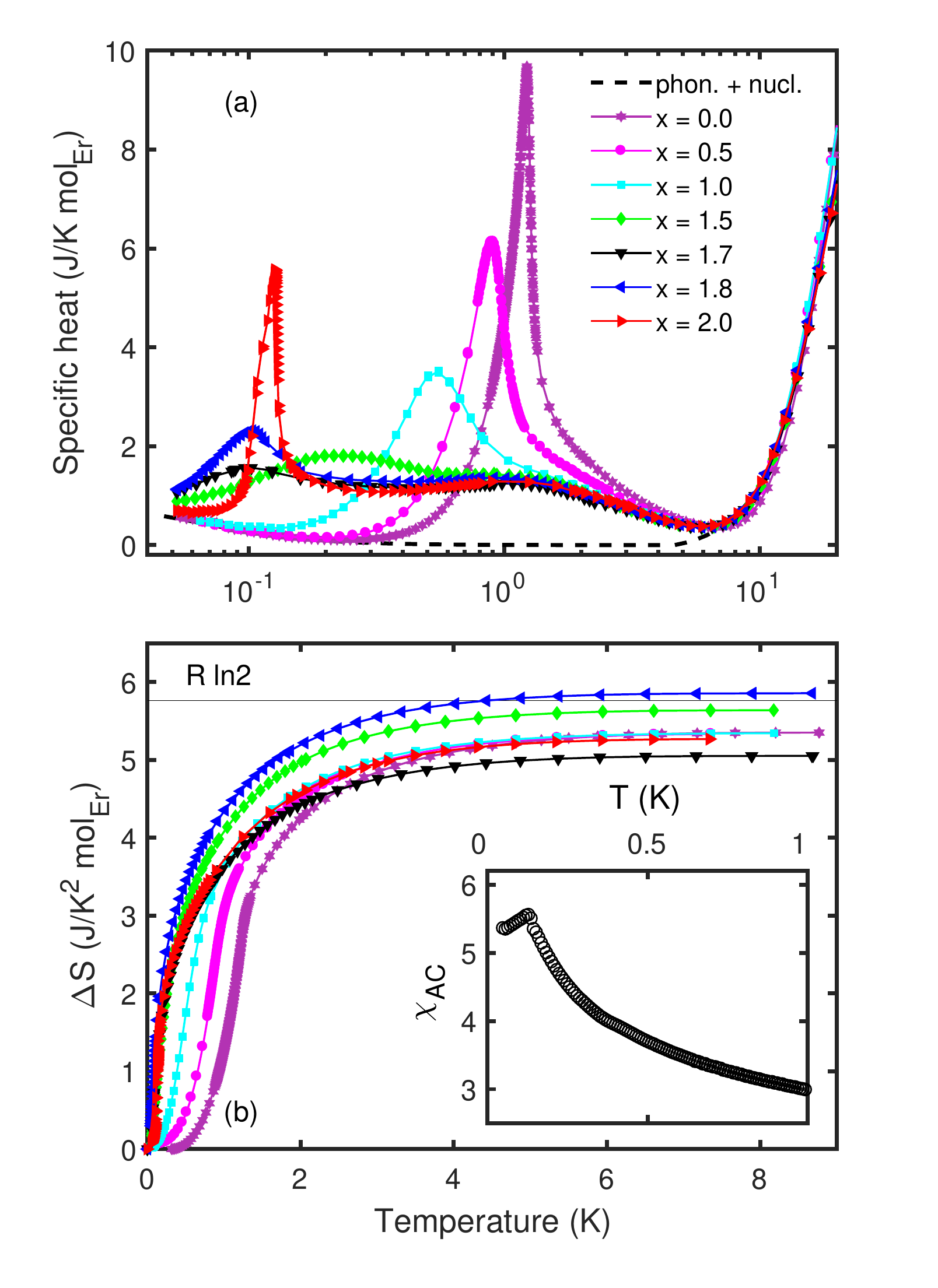}
	\caption{(Color online) (a) Temperature dependence of the total specific heat for the investigated compositions. The dashed line represents the lattice and nuclear contributions as explained in the text. (b) Integrated electronic magnetic entropy of the Er$_2$Ti$_{2-x}$Sn$_x$O$_7$ series per mole of Er$^{3+}$ ion. Theoretical level for a low-temperature doublet is indicates by a horizontal line. Inset shows the real component of the ac susceptibility for the $x = 2.0$ sample.}
	\label{fig-Cpseries}
\end{figure}

The evolution of thermodynamic behavior with doping has been monitored using heat capacity. In Fig.~\ref{fig-Cpseries} we present the temperature dependence of specific heat for all the compositions. At temperatures above $T \sim 6$ K all the curves are overlapping and specific heat can be well described with phononic contributions. At very low temperatures (below $\sim 0.1$ K) the nuclear contribution seems to dominate the measured specific heat for some compositions (notably $x = 0.5$ and $x = 1.0$). Figure~\ref{fig-Cpseries} shows the fits used to subtract both lattice and nuclear contributions (dashed line). These fits were obtained from scaling polynomial fits to the heat capacity measurements on the structurally similar non-magnetic material Lu$_2$Ti$_2$O$_7$ and the nuclear contribution calculated for Er$_2$Gd$_2$O$_7$ (see Supplemental Material of Ref.~\cite{Dun2015}). All compositions exhibit pronounced features at intermediate 
temperatures, which we associate with long-range magnetic ordering, also confirmed by neutron scattering results (see below). The transition temperature $T_N$ decreases linearly with $x$ and at the same time the peak becomes smaller. This trend is seen up to $x = 1.7$ where a small peak occurs just below 100 mK. For $x = 1.8$ we already notice a reversed trend with $T_N$ occurring above 100 mK and the peak height becomes visibly taller. Finally, for the end member of the series Er$_2$Sn$_2$O$_7$ we see a sharp peak positioned at 130 mK, revealing for the first time a LRO for this composition. The inset of the lower panel shows the ac magnetic susceptibility of the same composition exhibiting a sharp peak around the same temperature, corroborating the claim of LRO. We also notice a broad maximum centered just above 1 K, especially pronounced for higher $x$. Similar feature has been observed in several other pyrochlore systems~\cite{Dun2015,Yaouanc2013,Hallas2016} and it has been ascribed to exchange splitting of the ground state doublet due to the build up of magnetic correlations at low temperatures~\cite{Yaouanc2013}.

%
%

The electronic magnetic entropy associated with the low-lying crystal field levels of Er$^{3+}$ ions can be obtained integrating the specific heat after subtraction of the lattice and nuclear contributions. The results are presented in the lower panel of Fig.~\ref{fig-Cpseries} and show that the saturation level remains close to $R$ln 2 for $x = 1.5$ and $1.8$, an indicative of the presence of a Kramers doublet. The remaining compositions show a somewhat lower value of the integrated entropy, which might be related to systematic errors associated with the nuclear 
contribution subtraction. This is especially important in the case of $x = 0.5$, $1.0$ and $2.0$, in which case the nuclear magnetic contribution corresponds to a substantial part of the total measured specific heat at low temperatures.

\begin{figure*}
	\includegraphics[width=0.82\textwidth]{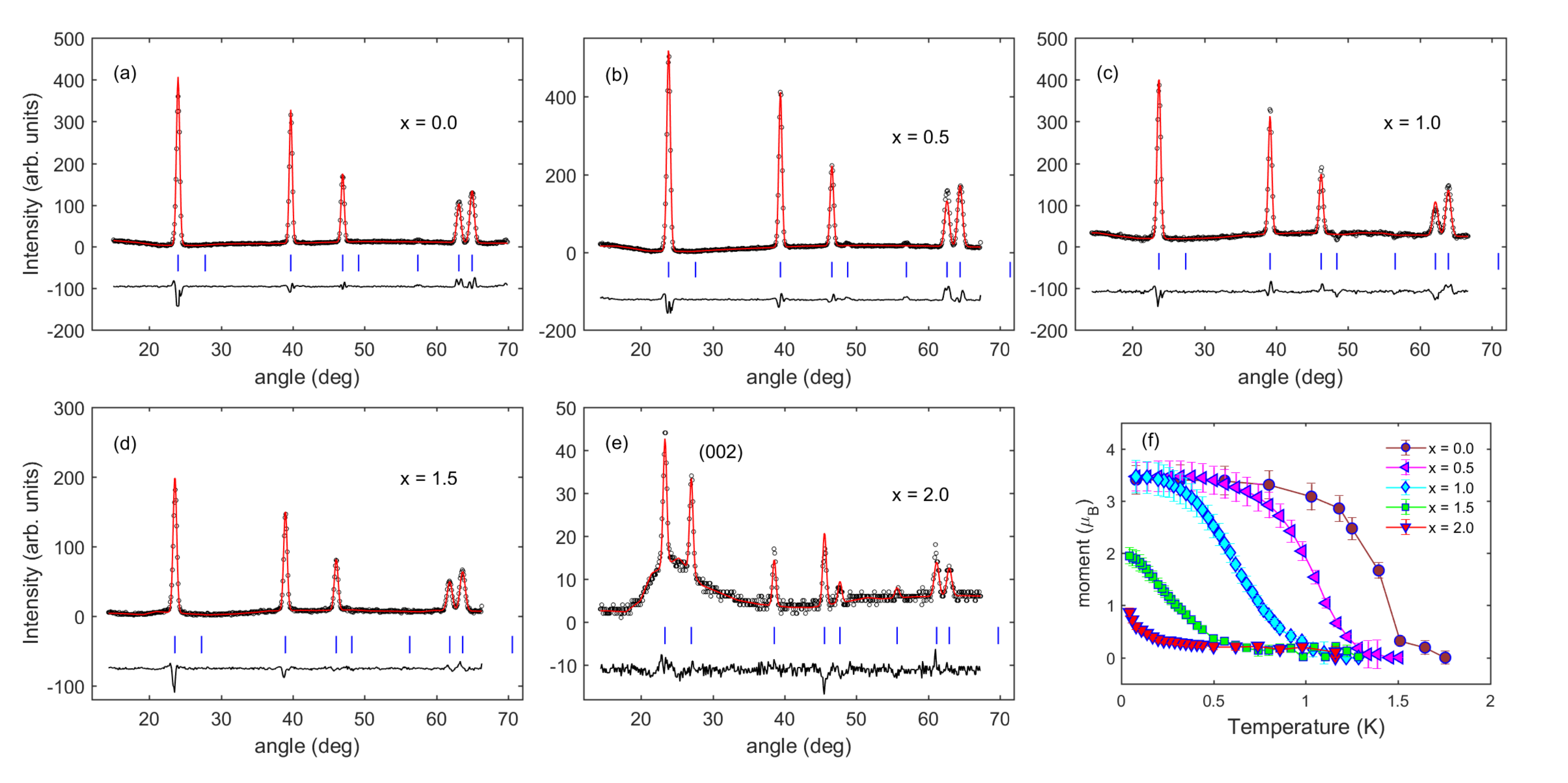}
	\caption{(Color online) (a) - (e) Magnetic diffraction profiles for the investigated compositions. The data is obtained by subtracting the profile above the ordering temperature ($\sim 1.3$ K) from the profile obtained within the ordered state ($\sim 100$ mK). The line below each graph is the difference between the magnetic diffraction profiles and the prediction of the model. The modeling of the diffuse background has not been attempted. (f) Temperature dependence of the ordered moment for each investigated composition obtained from the refinement of neutron scattering profiles.}
	\label{fig-neutrons}
\end{figure*}

To confirm the nature of ordered states, we present the results of neutron scattering experiments. In Fig.~\ref{fig-neutrons}a-f we plot the magnetic diffraction pattern for all the investigated compositions ($x = 0.0, 0.5, 1.0, 1.5, 2.0$). $x = 0$ exhibits a usual magnetic diffraction pattern with magnetic moments oriented according to the $\psi_2$ basis vector of the $\Gamma_5$ irreducible representation~\cite{Champion2003} (although $\psi_3$ cannot be excluded based on powder diffraction profile). Other compositions up to $x = 1.5$ show the same pattern, with the visible shift of diffraction peaks, reflecting the increase of the unit cell size (see inset of Fig.~\ref{fig-diffraction}). On the other hand, the pattern of $x = 2.0$ composition markedly differs from others, with a new magnetic Bragg peak at [002]. The diffraction profile can be successfully described using the Palmer-Chalker model~\cite{Palmer2000} of the $\Gamma_7$ irreducible representation, with the ordering vector \textbf{k} = 0. It is noteworthy to mention that we also observe a broad, diffuse scattering as reported before~\cite{Sarte2011,Guitteny2013}. Since it has been argued that this diffuse scattering originates from the same $\Gamma_7$ representation~\cite{Guitteny2013}, it leads to the conclusion that our Er$_2$Sn$_2$O$_7$ sample exhibits simultaneously long- and short-range order effects. This could be attributed to local imperfections in the crystal structure, not seen in our x-ray and neutron profiles, resulting in short-range correlated islands of spins not connected to the majority of ordered phase. Similar coexistence of long- and short-range order effects have been observed in Yb$_2$Sn$_2$O$_7$~\cite{Lago2014}.

The temperature dependence of the refined moment for different compositions is presented in Fig.~\ref{fig-neutrons}f. Three lower $x$ compounds saturate around the same value at low temperatures, $\mu_{sat} \approx 3.45 \mu_B$. On the other hand, $x = 1.5$ compound shows a reduced moment and would reach a maximum of around $2 \mu_B$, although the complete saturation has not been observed. Even more drastic case is presented for $x = 2.0$ where the refined moment shows an upward trend even at the lowest temperature (70 mK) and reaches only $\sim 0.9 \mu_B$. In any case, it is safe to conclude that the observed moment is significantly reduced compared to the Ti analog.

%
%

\begin{figure}[bh!]
	\includegraphics[width=0.35\textwidth]{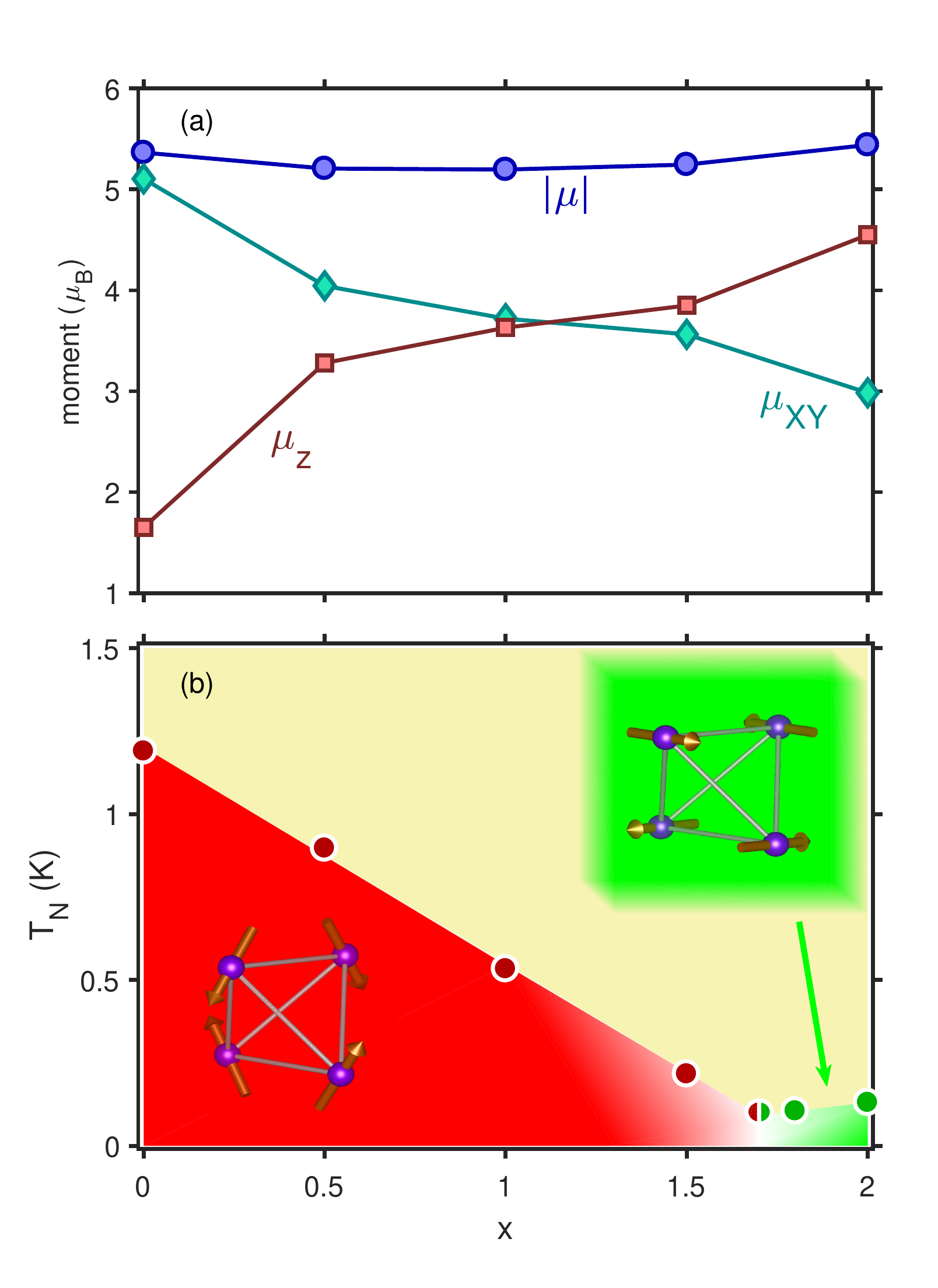}
	\caption{(Color online) (a) The variation of the single-ion anisotropy of erbium moments with doping. $\mu_{XY}$ and $\mu_z$ designate components perpendicular and parallel to the $<$111$>$ axis, respectively. $|\mu|$ is the total moment value. (b) Phase diagram of the Er$_2$Ti$_{2-x}$Sn$_x$O$_7$ series. The $x = 1.7$ composition should be close to the expected quantum phase transition between $\psi_2$ and Palmer-Chalker phases. The exact shape of phase boundaries in that region is unknown and it might be strongly influenced by exact processes of sample preparation.}
	\label{fig-moment}
\end{figure}

To address the question of the reduction of the magnetic moment for $x = 1.5$ and $x = 2.0$ it is instructive to take into account the doping dependence of the single-ion anisotropy of the erbium ion. Inelastic neutron scattering experiments have been performed in order to obtain the crystal field level scheme for each compound. The corresponding wavefunctions were then calculated using the refined values of Stevens' operators. In order to calculate the magnetic moments, only the ground state Kramers doublets have been taken into account. Figure~\ref{fig-moment}a displays the results of the crystal field level analysis~\cite{Masae2007} which indicates that the anisotropy gradually shifts from almost XY-like in the $x = 0$ compound to more Ising-like for $x = 2.0$, with the moment size unchanged across the doping level. The cross-over occurs between $x = 1.0$ and $x = 1.5$, correlating with the reduction of the value of the ordered moment at low temperature. We point out that the co-existence of ordered and fluctuating components of the magnetic moment have been recently demonstrated in Nd$_2$Zr$_2$O$_7$ pyrochlore~\cite{Petit2016}, based on the theoretical concept of magnetic moment fragmentation~\cite{BrooksBartlett2014}. At low temperatures the Nd-based system exhibits magnetic Bragg peaks, characteristic of LRO, while at the same time displaying pinch points in its inelastic scattering profile, indicative of the dynamical Coulomb phase. It is important to emphasize that both of these components, ordered and fluctuating, are oriented along the $<$111$>$ axis. In Er$_2$Sn$_2$O$_7$, to the contrary, two components are orthogonal to each other. 

%
%

In Fig.~\ref{fig-moment}b we plot how the ordering temperature obtained from the heat capacity measurements depends on $x$. For $0 \leq x \leq 1.5$ a linear decrease of $T_N$ is observed. Although $x = 1.7$ composition seems to continue along the same trend, it is difficult to argue that it belongs to the same ordering type ($\psi_2$) because the same argument could be used when looking at the trend of $x = 1.8$ and $x = 2.0$ compositions (Palmer-Chalker type). The most reasonable assumption is that around $x = 1.7$ there is a cross-over from one type of ordering to another which in real samples is heavily influenced by extreme sensitivity to local inhomogeneities caused by the process of sample preparation. This goes in line with the observed diffuse scattering profile for $x = 2.0$. On the other hand, one cannot rule out that for ideally prepared samples there is a true quantum phase transition where the reduction of the moment is caused by incipient quantum fluctuations. Indeed, recent theoretical consideration~\cite{Yan2017} of the phase boundary between $\psi_2$ and Palmer-Chalker configurations claims that the ordered moment within the $\psi_2$ state is strongly renormalized, with logarithmically divergent corrections as the boundary is approached from the $x = 0.0$ side. On the other hand, they suggest that on the Palmer-Chalker side of the boundary the renormalization is rather small, at the order of 10\%. The contradiction with our observations is not surprising knowing that the spin-wave theory underestimates effects of quantum fluctuations. An alternative scenario, based on extensive Monte Carlo simulations~\cite{Zhitomirsky2014}, indicate a possibility that a $\psi_2$ -- Palmer-Chalker boundary is characterized by a non-zero ordering temperature. Nevertheless, due to proven difficulties of preparing 'ideal' stannate samples, future progress in this direction will come only after painstaking accumulation of many studies. We hope that our study will provide a good starting point for more thorough understanding of the processes underlying this hotly debated topic.

%
%
%
%


%
%
%
%
%

To conclude, we have presented the evolution of long-range order in the series of samples Er$_2$Ti$_{2-x}$Sn$_x$O$_7$. The end member of the series, Er$_2$Sn$_2$O$_7$, does order below 130 mK in the Palmer-Chalker configuration, at the same time showing short-range correlation effects in the form of diffuse scattering. The details of the cross-over between the $\psi_2$ state and the Palmer-Chalker state are expected to be strongly influenced by the sample preparation. Given that XY-anisotropy of long-range order is preserved across the series, and that for ideally prepared samples around $x = 1.7$ one could expect to have a true quantum phase transition, this system would represent a rare example of an XY-dominated quantum spin-liquid, which from a theoretical point of view has been seldom addressed.


%
%
%
%
%

%

RSF would like to acknowledge support from the Brazilian agencies CNPq (306614/2015-4) and FAPESP (2015/16191-5).


\begin{thebibliography}{35}
	\expandafter\ifx\csname natexlab\endcsname\relax\def\natexlab#1{#1}\fi
	\expandafter\ifx\csname bibnamefont\endcsname\relax
	\def\bibnamefont#1{#1}\fi
	\expandafter\ifx\csname bibfnamefont\endcsname\relax
	\def\bibfnamefont#1{#1}\fi
	\expandafter\ifx\csname citenamefont\endcsname\relax
	\def\citenamefont#1{#1}\fi
	\expandafter\ifx\csname url\endcsname\relax
	\def\url#1{\texttt{#1}}\fi
	\expandafter\ifx\csname urlprefix\endcsname\relax\def\urlprefix{URL }\fi
	\providecommand{\bibinfo}[2]{#2}
	\providecommand{\eprint}[2][]{\url{#2}}
	
	\bibitem[{\citenamefont{Fennell et~al.}(2009)\citenamefont{Fennell, Deen,
			Wildes, Schmalzl, Prabhakaran, Boothroyd, Aldus, McMorrow, and
			Bramwell}}]{Fennell2009}
	\bibinfo{author}{\bibfnamefont{T.}~\bibnamefont{Fennell}},
	\bibinfo{author}{\bibfnamefont{P.~P.} \bibnamefont{Deen}},
	\bibinfo{author}{\bibfnamefont{A.~R.} \bibnamefont{Wildes}},
	\bibinfo{author}{\bibfnamefont{K.}~\bibnamefont{Schmalzl}},
	\bibinfo{author}{\bibfnamefont{D.}~\bibnamefont{Prabhakaran}},
	\bibinfo{author}{\bibfnamefont{A.~T.} \bibnamefont{Boothroyd}},
	\bibinfo{author}{\bibfnamefont{R.~J.} \bibnamefont{Aldus}},
	\bibinfo{author}{\bibfnamefont{D.~F.} \bibnamefont{McMorrow}},
	\bibnamefont{and} \bibinfo{author}{\bibfnamefont{S.~T.}
		\bibnamefont{Bramwell}}, \bibinfo{journal}{Science}
	\textbf{\bibinfo{volume}{326}}, \bibinfo{pages}{415} (\bibinfo{year}{2009}).
	
	\bibitem[{\citenamefont{Ramirez et~al.}(1999)\citenamefont{Ramirez, Hayashi,
			Cava, Siddharthan, and Shastry}}]{Ramirez1999}
	\bibinfo{author}{\bibfnamefont{A.~P.} \bibnamefont{Ramirez}},
	\bibinfo{author}{\bibfnamefont{A.}~\bibnamefont{Hayashi}},
	\bibinfo{author}{\bibfnamefont{R.~J.} \bibnamefont{Cava}},
	\bibinfo{author}{\bibfnamefont{R.}~\bibnamefont{Siddharthan}},
	\bibnamefont{and} \bibinfo{author}{\bibfnamefont{B.~S.}
		\bibnamefont{Shastry}}, \bibinfo{journal}{Nature}
	\textbf{\bibinfo{volume}{399}}, \bibinfo{pages}{333} (\bibinfo{year}{1999}).
	
	\bibitem[{\citenamefont{Lago et~al.}(2010)\citenamefont{Lago, \v{Z}ivkovi\'{c},
			Malkin, Fernandez, Ghigna, de~Reotier, Yaouanc, and Rojo}}]{Lago2010}
	\bibinfo{author}{\bibfnamefont{J.}~\bibnamefont{Lago}},
	\bibinfo{author}{\bibfnamefont{I.}~\bibnamefont{\v{Z}ivkovi\'{c}}},
	\bibinfo{author}{\bibfnamefont{B.~Z.} \bibnamefont{Malkin}},
	\bibinfo{author}{\bibfnamefont{J.~R.} \bibnamefont{Fernandez}},
	\bibinfo{author}{\bibfnamefont{P.}~\bibnamefont{Ghigna}},
	\bibinfo{author}{\bibfnamefont{P.~D.} \bibnamefont{de~Reotier}},
	\bibinfo{author}{\bibfnamefont{A.}~\bibnamefont{Yaouanc}}, \bibnamefont{and}
	\bibinfo{author}{\bibfnamefont{T.}~\bibnamefont{Rojo}},
	\bibinfo{journal}{Phys.Rev.Lett.} \textbf{\bibinfo{volume}{104}},
	\bibinfo{pages}{247203} (\bibinfo{year}{2010}).
	
	\bibitem[{\citenamefont{de~R\'{e}otier
			et~al.}(2012)\citenamefont{de~R\'{e}otier, Yaouanc, Chapuis, Curnoe, Grenier,
			Ressouchea, Marin, Lago, Baines, and Giblin}}]{Reotier2012}
	\bibinfo{author}{\bibfnamefont{P.~D.} \bibnamefont{de~R\'{e}otier}},
	\bibinfo{author}{\bibfnamefont{A.}~\bibnamefont{Yaouanc}},
	\bibinfo{author}{\bibfnamefont{Y.}~\bibnamefont{Chapuis}},
	\bibinfo{author}{\bibfnamefont{S.~H.} \bibnamefont{Curnoe}},
	\bibinfo{author}{\bibfnamefont{B.}~\bibnamefont{Grenier}},
	\bibinfo{author}{\bibfnamefont{E.}~\bibnamefont{Ressouchea}},
	\bibinfo{author}{\bibfnamefont{C.}~\bibnamefont{Marin}},
	\bibinfo{author}{\bibfnamefont{J.}~\bibnamefont{Lago}},
	\bibinfo{author}{\bibfnamefont{C.}~\bibnamefont{Baines}}, \bibnamefont{and}
	\bibinfo{author}{\bibfnamefont{S.~R.} \bibnamefont{Giblin}},
	\bibinfo{journal}{Phys.Rev.B} \textbf{\bibinfo{volume}{86}},
	\bibinfo{pages}{104424} (\bibinfo{year}{2012}).
	
	\bibitem[{\citenamefont{Poole et~al.}(2007)\citenamefont{Poole, Wills, and
			Lelievre-Berna}}]{Poole2007}
	\bibinfo{author}{\bibfnamefont{A.}~\bibnamefont{Poole}},
	\bibinfo{author}{\bibfnamefont{A.~S.} \bibnamefont{Wills}}, \bibnamefont{and}
	\bibinfo{author}{\bibfnamefont{E.}~\bibnamefont{Lelievre-Berna}},
	\bibinfo{journal}{J.Phys.:Condens.Matter} \textbf{\bibinfo{volume}{19}},
	\bibinfo{pages}{452201} (\bibinfo{year}{2007}).
	
	\bibitem[{\citenamefont{Guitteny et~al.}(2013)\citenamefont{Guitteny, Petit,
			nad J.~Robert, Bonville, Forget, and Mirebeau}}]{Guitteny2013}
	\bibinfo{author}{\bibfnamefont{S.}~\bibnamefont{Guitteny}},
	\bibinfo{author}{\bibfnamefont{S.}~\bibnamefont{Petit}},
	\bibinfo{author}{\bibfnamefont{E.~L.} \bibnamefont{nad J.~Robert}},
	\bibinfo{author}{\bibfnamefont{P.}~\bibnamefont{Bonville}},
	\bibinfo{author}{\bibfnamefont{A.}~\bibnamefont{Forget}}, \bibnamefont{and}
	\bibinfo{author}{\bibfnamefont{I.}~\bibnamefont{Mirebeau}},
	\bibinfo{journal}{Phys.Rev.B} \textbf{\bibinfo{volume}{88}},
	\bibinfo{pages}{134408} (\bibinfo{year}{2013}).
	
	\bibitem[{\citenamefont{Villain et~al.}(1980)\citenamefont{Villain, Bidaux,
			Carton, and Conte}}]{Villain1980}
	\bibinfo{author}{\bibfnamefont{J.}~\bibnamefont{Villain}},
	\bibinfo{author}{\bibfnamefont{R.}~\bibnamefont{Bidaux}},
	\bibinfo{author}{\bibfnamefont{J.-P.} \bibnamefont{Carton}},
	\bibnamefont{and} \bibinfo{author}{\bibfnamefont{R.}~\bibnamefont{Conte}},
	\bibinfo{journal}{J.Phys. (France)} \textbf{\bibinfo{volume}{41}},
	\bibinfo{pages}{1263} (\bibinfo{year}{1980}).
	
	\bibitem[{\citenamefont{Champion et~al.}(2003)\citenamefont{Champion, Harris,
			Holdsworth, Wills, Balakrishnan, Bramwell, \v{C}i\v{z}m\'{a}r, Fennell,
			Gardner, Lago et~al.}}]{Champion2003}
	\bibinfo{author}{\bibfnamefont{J.~D.~M.} \bibnamefont{Champion}},
	\bibinfo{author}{\bibfnamefont{M.~J.} \bibnamefont{Harris}},
	\bibinfo{author}{\bibfnamefont{P.~C.~W.} \bibnamefont{Holdsworth}},
	\bibinfo{author}{\bibfnamefont{A.~S.} \bibnamefont{Wills}},
	\bibinfo{author}{\bibfnamefont{G.}~\bibnamefont{Balakrishnan}},
	\bibinfo{author}{\bibfnamefont{S.~T.} \bibnamefont{Bramwell}},
	\bibinfo{author}{\bibfnamefont{E.}~\bibnamefont{\v{C}i\v{z}m\'{a}r}},
	\bibinfo{author}{\bibfnamefont{T.}~\bibnamefont{Fennell}},
	\bibinfo{author}{\bibfnamefont{J.~S.} \bibnamefont{Gardner}},
	\bibinfo{author}{\bibfnamefont{J.}~\bibnamefont{Lago}}, \bibnamefont{et~al.},
	\bibinfo{journal}{Phys.Rev.B} \textbf{\bibinfo{volume}{68}},
	\bibinfo{pages}{020401(R)} (\bibinfo{year}{2003}).
	
	\bibitem[{\citenamefont{Savary et~al.}(2012)\citenamefont{Savary, Ross, Gaulin,
			Ruff, and Balents}}]{Savary2012}
	\bibinfo{author}{\bibfnamefont{L.}~\bibnamefont{Savary}},
	\bibinfo{author}{\bibfnamefont{K.~A.} \bibnamefont{Ross}},
	\bibinfo{author}{\bibfnamefont{B.~D.} \bibnamefont{Gaulin}},
	\bibinfo{author}{\bibfnamefont{J.~P.~C.} \bibnamefont{Ruff}},
	\bibnamefont{and} \bibinfo{author}{\bibfnamefont{L.}~\bibnamefont{Balents}},
	\bibinfo{journal}{Phys.Rev.Lett.} \textbf{\bibinfo{volume}{109}},
	\bibinfo{pages}{167201} (\bibinfo{year}{2012}).
	
	\bibitem[{\citenamefont{Zhitomirsky et~al.}(2012)\citenamefont{Zhitomirsky,
			Gvozdikova, Holdsworth, and Moessner}}]{Zhitomirsky2012}
	\bibinfo{author}{\bibfnamefont{M.~E.} \bibnamefont{Zhitomirsky}},
	\bibinfo{author}{\bibfnamefont{M.~V.} \bibnamefont{Gvozdikova}},
	\bibinfo{author}{\bibfnamefont{P.~C.~W.} \bibnamefont{Holdsworth}},
	\bibnamefont{and} \bibinfo{author}{\bibfnamefont{R.}~\bibnamefont{Moessner}},
	\bibinfo{journal}{Phys.Rev.Lett.} \textbf{\bibinfo{volume}{109}},
	\bibinfo{pages}{077204} (\bibinfo{year}{2012}).
	
	\bibitem[{\citenamefont{Niven et~al.}(2014)\citenamefont{Niven, Johnson,
			Bourque, Murray, James, Dabkowska, Gaulin, and White}}]{Niven2014}
	\bibinfo{author}{\bibfnamefont{J.~F.} \bibnamefont{Niven}},
	\bibinfo{author}{\bibfnamefont{M.~B.} \bibnamefont{Johnson}},
	\bibinfo{author}{\bibfnamefont{A.}~\bibnamefont{Bourque}},
	\bibinfo{author}{\bibfnamefont{P.~J.} \bibnamefont{Murray}},
	\bibinfo{author}{\bibfnamefont{D.~D.} \bibnamefont{James}},
	\bibinfo{author}{\bibfnamefont{H.~A.} \bibnamefont{Dabkowska}},
	\bibinfo{author}{\bibfnamefont{B.~D.} \bibnamefont{Gaulin}},
	\bibnamefont{and} \bibinfo{author}{\bibfnamefont{M.~A.} \bibnamefont{White}},
	\bibinfo{journal}{Proc.R.Soc.London A} \textbf{\bibinfo{volume}{470}},
	\bibinfo{pages}{20140387} (\bibinfo{year}{2014}).
	
	\bibitem[{\citenamefont{Maryasin and Zhitomirsky}(2014)}]{Maryasin2014}
	\bibinfo{author}{\bibfnamefont{V.~S.} \bibnamefont{Maryasin}} \bibnamefont{and}
	\bibinfo{author}{\bibfnamefont{M.~E.} \bibnamefont{Zhitomirsky}},
	\bibinfo{journal}{Phys.Rev.B} \textbf{\bibinfo{volume}{90}},
	\bibinfo{pages}{094412} (\bibinfo{year}{2014}).
	
	\bibitem[{\citenamefont{Andreanov and McClarty}(2015)}]{Adreanov2015}
	\bibinfo{author}{\bibfnamefont{A.}~\bibnamefont{Andreanov}} \bibnamefont{and}
	\bibinfo{author}{\bibfnamefont{P.~A.} \bibnamefont{McClarty}},
	\bibinfo{journal}{Phys.Rev.B} \textbf{\bibinfo{volume}{91}},
	\bibinfo{pages}{064401} (\bibinfo{year}{2015}).
	
	\bibitem[{\citenamefont{Gaudet et~al.}(2016)\citenamefont{Gaudet, Hallas,
			Maharaj, Buhariwalla, Kermarrec, Butch, Munsie, Dabkowska, Luke, and
			Gauli}}]{Gaudet2016}
	\bibinfo{author}{\bibfnamefont{J.}~\bibnamefont{Gaudet}},
	\bibinfo{author}{\bibfnamefont{A.~M.} \bibnamefont{Hallas}},
	\bibinfo{author}{\bibfnamefont{D.~D.} \bibnamefont{Maharaj}},
	\bibinfo{author}{\bibfnamefont{C.~R.~C.} \bibnamefont{Buhariwalla}},
	\bibinfo{author}{\bibfnamefont{E.}~\bibnamefont{Kermarrec}},
	\bibinfo{author}{\bibfnamefont{N.~P.} \bibnamefont{Butch}},
	\bibinfo{author}{\bibfnamefont{T.~J.~S.} \bibnamefont{Munsie}},
	\bibinfo{author}{\bibfnamefont{H.~A.} \bibnamefont{Dabkowska}},
	\bibinfo{author}{\bibfnamefont{G.~M.} \bibnamefont{Luke}}, \bibnamefont{and}
	\bibinfo{author}{\bibfnamefont{B.~D.} \bibnamefont{Gauli}},
	\bibinfo{journal}{Phys.Rev.B} \textbf{\bibinfo{volume}{94}},
	\bibinfo{pages}{060407(R)} (\bibinfo{year}{2016}).
	
	\bibitem[{\citenamefont{Wong et~al.}(2013)\citenamefont{Wong, Hao, and
			Gingras}}]{Wong2013}
	\bibinfo{author}{\bibfnamefont{A.~W.~C.} \bibnamefont{Wong}},
	\bibinfo{author}{\bibfnamefont{Z.}~\bibnamefont{Hao}}, \bibnamefont{and}
	\bibinfo{author}{\bibfnamefont{M.~J.~P.} \bibnamefont{Gingras}},
	\bibinfo{journal}{Phys.Rev.B} \textbf{\bibinfo{volume}{88}},
	\bibinfo{pages}{144402} (\bibinfo{year}{2013}).
	
	\bibitem[{\citenamefont{Dun et~al.}(2015)\citenamefont{Dun, Li, Freitas,
			Arrighi, Cruz, Lee, Choi, Cao, Silverstein, Wiebe et~al.}}]{Dun2015}
	\bibinfo{author}{\bibfnamefont{Z.~L.} \bibnamefont{Dun}},
	\bibinfo{author}{\bibfnamefont{X.}~\bibnamefont{Li}},
	\bibinfo{author}{\bibfnamefont{R.~S.} \bibnamefont{Freitas}},
	\bibinfo{author}{\bibfnamefont{E.}~\bibnamefont{Arrighi}},
	\bibinfo{author}{\bibfnamefont{C.~R.~D.} \bibnamefont{Cruz}},
	\bibinfo{author}{\bibfnamefont{M.}~\bibnamefont{Lee}},
	\bibinfo{author}{\bibfnamefont{E.~S.} \bibnamefont{Choi}},
	\bibinfo{author}{\bibfnamefont{H.~B.} \bibnamefont{Cao}},
	\bibinfo{author}{\bibfnamefont{H.~J.} \bibnamefont{Silverstein}},
	\bibinfo{author}{\bibfnamefont{C.~R.} \bibnamefont{Wiebe}},
	\bibnamefont{et~al.}, \bibinfo{journal}{Phys.Rev.B}
	\textbf{\bibinfo{volume}{92}}, \bibinfo{pages}{140407(R)}
	(\bibinfo{year}{2015}).
	
	\bibitem[{\citenamefont{Matsuhira et~al.}(2004)\citenamefont{Matsuhira,
			Hinatsu, Tenya, Amitsuka, and Sakakibara}}]{Matsuhira2004}
	\bibinfo{author}{\bibfnamefont{K.}~\bibnamefont{Matsuhira}},
	\bibinfo{author}{\bibfnamefont{Y.}~\bibnamefont{Hinatsu}},
	\bibinfo{author}{\bibfnamefont{K.}~\bibnamefont{Tenya}},
	\bibinfo{author}{\bibfnamefont{H.}~\bibnamefont{Amitsuka}}, \bibnamefont{and}
	\bibinfo{author}{\bibfnamefont{T.}~\bibnamefont{Sakakibara}},
	\bibinfo{journal}{J.Phys.Soc.Jpn} \textbf{\bibinfo{volume}{71}},
	\bibinfo{pages}{1576} (\bibinfo{year}{2004}).
	
	\bibitem[{\citenamefont{Gardner et~al.}(1999)\citenamefont{Gardner, Dunsiger,
			Gaulin, Gingras, Greedan, Kieﬂ, Lumsden, MacFarlane, Raju, Sonier
			et~al.}}]{Gardner1999}
	\bibinfo{author}{\bibfnamefont{J.~S.} \bibnamefont{Gardner}},
	\bibinfo{author}{\bibfnamefont{S.~R.} \bibnamefont{Dunsiger}},
	\bibinfo{author}{\bibfnamefont{B.~D.} \bibnamefont{Gaulin}},
	\bibinfo{author}{\bibfnamefont{M.~J.~P.} \bibnamefont{Gingras}},
	\bibinfo{author}{\bibfnamefont{J.~E.} \bibnamefont{Greedan}},
	\bibinfo{author}{\bibfnamefont{R.~F.} \bibnamefont{Kieﬂ}},
	\bibinfo{author}{\bibfnamefont{M.~D.} \bibnamefont{Lumsden}},
	\bibinfo{author}{\bibfnamefont{W.~A.} \bibnamefont{MacFarlane}},
	\bibinfo{author}{\bibfnamefont{N.~P.} \bibnamefont{Raju}},
	\bibinfo{author}{\bibfnamefont{J.~E.} \bibnamefont{Sonier}},
	\bibnamefont{et~al.}, \bibinfo{journal}{Phys.Rev.Lett.}
	\textbf{\bibinfo{volume}{82}}, \bibinfo{pages}{1012} (\bibinfo{year}{1999}).
	
	\bibitem[{\citenamefont{Mirebeau et~al.}(2005)\citenamefont{Mirebeau, Apetrei,
			nad P.~Bonville, Forget, Colson, Glazkov, Sanchez, Isnard, and
			Suard}}]{Mirebeau2005}
	\bibinfo{author}{\bibfnamefont{I.}~\bibnamefont{Mirebeau}},
	\bibinfo{author}{\bibfnamefont{A.}~\bibnamefont{Apetrei}},
	\bibinfo{author}{\bibfnamefont{J.~R.-C.} \bibnamefont{nad P.~Bonville}},
	\bibinfo{author}{\bibfnamefont{A.}~\bibnamefont{Forget}},
	\bibinfo{author}{\bibfnamefont{D.}~\bibnamefont{Colson}},
	\bibinfo{author}{\bibfnamefont{V.}~\bibnamefont{Glazkov}},
	\bibinfo{author}{\bibfnamefont{J.~P.} \bibnamefont{Sanchez}},
	\bibinfo{author}{\bibfnamefont{O.}~\bibnamefont{Isnard}}, \bibnamefont{and}
	\bibinfo{author}{\bibfnamefont{E.}~\bibnamefont{Suard}},
	\bibinfo{journal}{Phys.Rev.Lett.} \textbf{\bibinfo{volume}{94}},
	\bibinfo{pages}{246402} (\bibinfo{year}{2005}).
	
	\bibitem[{\citenamefont{Dahlberg et~al.}(2011)\citenamefont{Dahlberg, Matthews,
			Jiramongkolchai, Cava, and Schiffer}}]{Dahlberg2001}
	\bibinfo{author}{\bibfnamefont{M.~L.} \bibnamefont{Dahlberg}},
	\bibinfo{author}{\bibfnamefont{M.~J.} \bibnamefont{Matthews}},
	\bibinfo{author}{\bibfnamefont{P.}~\bibnamefont{Jiramongkolchai}},
	\bibinfo{author}{\bibfnamefont{R.~J.} \bibnamefont{Cava}}, \bibnamefont{and}
	\bibinfo{author}{\bibfnamefont{P.}~\bibnamefont{Schiffer}},
	\bibinfo{journal}{Phys.Rev.B} \textbf{\bibinfo{volume}{83}},
	\bibinfo{pages}{140410(R)} (\bibinfo{year}{2011}).
	
	\bibitem[{\citenamefont{Chang et~al.}(2012)\citenamefont{Chang, Onoda, Su, Kao,
			Tsuei, Yasui, Kakurai, and Lees}}]{Chang2012}
	\bibinfo{author}{\bibfnamefont{L.-J.} \bibnamefont{Chang}},
	\bibinfo{author}{\bibfnamefont{S.}~\bibnamefont{Onoda}},
	\bibinfo{author}{\bibfnamefont{Y.}~\bibnamefont{Su}},
	\bibinfo{author}{\bibfnamefont{Y.-J.} \bibnamefont{Kao}},
	\bibinfo{author}{\bibfnamefont{K.-D.} \bibnamefont{Tsuei}},
	\bibinfo{author}{\bibfnamefont{Y.}~\bibnamefont{Yasui}},
	\bibinfo{author}{\bibfnamefont{K.}~\bibnamefont{Kakurai}}, \bibnamefont{and}
	\bibinfo{author}{\bibfnamefont{M.~R.} \bibnamefont{Lees}},
	\bibinfo{journal}{Nat. Commun.} \textbf{\bibinfo{volume}{3}},
	\bibinfo{pages}{992} (\bibinfo{year}{2012}).
	
	\bibitem[{\citenamefont{Ross et~al.}(2012)\citenamefont{Ross, Proffen,
			Dabkowska, Quilliam, Yaraskavitch, Kycia, and Gaulin}}]{Ross2012}
	\bibinfo{author}{\bibfnamefont{K.~A.} \bibnamefont{Ross}},
	\bibinfo{author}{\bibfnamefont{T.}~\bibnamefont{Proffen}},
	\bibinfo{author}{\bibfnamefont{H.~A.} \bibnamefont{Dabkowska}},
	\bibinfo{author}{\bibfnamefont{J.~A.} \bibnamefont{Quilliam}},
	\bibinfo{author}{\bibfnamefont{L.~R.} \bibnamefont{Yaraskavitch}},
	\bibinfo{author}{\bibfnamefont{J.~B.} \bibnamefont{Kycia}}, \bibnamefont{and}
	\bibinfo{author}{\bibfnamefont{B.~D.} \bibnamefont{Gaulin}},
	\bibinfo{journal}{Phys.Rev.B} \textbf{\bibinfo{volume}{86}},
	\bibinfo{pages}{174424} (\bibinfo{year}{2012}).
	
	\bibitem[{\citenamefont{Yaouanc et~al.}(2013)\citenamefont{Yaouanc,
			de~Re\'{o}tier, Bonville, Hodges, Glazkov, Keller, Sikolenko, Bartkowiak,
			Amato, Baines et~al.}}]{Yaouanc2013}
	\bibinfo{author}{\bibfnamefont{A.}~\bibnamefont{Yaouanc}},
	\bibinfo{author}{\bibfnamefont{P.~D.} \bibnamefont{de~Re\'{o}tier}},
	\bibinfo{author}{\bibfnamefont{P.}~\bibnamefont{Bonville}},
	\bibinfo{author}{\bibfnamefont{J.~A.} \bibnamefont{Hodges}},
	\bibinfo{author}{\bibfnamefont{V.}~\bibnamefont{Glazkov}},
	\bibinfo{author}{\bibfnamefont{L.}~\bibnamefont{Keller}},
	\bibinfo{author}{\bibfnamefont{V.}~\bibnamefont{Sikolenko}},
	\bibinfo{author}{\bibfnamefont{M.}~\bibnamefont{Bartkowiak}},
	\bibinfo{author}{\bibfnamefont{A.}~\bibnamefont{Amato}},
	\bibinfo{author}{\bibfnamefont{C.}~\bibnamefont{Baines}},
	\bibnamefont{et~al.}, \bibinfo{journal}{Phys.Rev.Lett.}
	\textbf{\bibinfo{volume}{110}}, \bibinfo{pages}{127207}
	(\bibinfo{year}{2013}).
	
	\bibitem[{\citenamefont{Lago et~al.}(2014)\citenamefont{Lago, \v{Z}ivkovi\'{c},
			Piatek, \'{A}lvarez, H\"{u}vonen, Pratt, D\'{i}az, and Rojo}}]{Lago2014}
	\bibinfo{author}{\bibfnamefont{J.}~\bibnamefont{Lago}},
	\bibinfo{author}{\bibfnamefont{I.}~\bibnamefont{\v{Z}ivkovi\'{c}}},
	\bibinfo{author}{\bibfnamefont{J.~O.} \bibnamefont{Piatek}},
	\bibinfo{author}{\bibfnamefont{P.}~\bibnamefont{\'{A}lvarez}},
	\bibinfo{author}{\bibfnamefont{D.}~\bibnamefont{H\"{u}vonen}},
	\bibinfo{author}{\bibfnamefont{F.~L.} \bibnamefont{Pratt}},
	\bibinfo{author}{\bibfnamefont{M.}~\bibnamefont{D\'{i}az}}, \bibnamefont{and}
	\bibinfo{author}{\bibfnamefont{T.}~\bibnamefont{Rojo}},
	\bibinfo{journal}{Phys.Rev.B} \textbf{\bibinfo{volume}{89}},
	\bibinfo{pages}{024421} (\bibinfo{year}{2014}).
	
	\bibitem[{\citenamefont{Lago et~al.}(2005)\citenamefont{Lago, Lancaster,
			Blundell, Bramwell, Pratt, Shirai, and Baines}}]{Lago2005}
	\bibinfo{author}{\bibfnamefont{J.}~\bibnamefont{Lago}},
	\bibinfo{author}{\bibfnamefont{T.}~\bibnamefont{Lancaster}},
	\bibinfo{author}{\bibfnamefont{S.~J.} \bibnamefont{Blundell}},
	\bibinfo{author}{\bibfnamefont{S.~T.} \bibnamefont{Bramwell}},
	\bibinfo{author}{\bibfnamefont{F.~L.} \bibnamefont{Pratt}},
	\bibinfo{author}{\bibfnamefont{M.}~\bibnamefont{Shirai}}, \bibnamefont{and}
	\bibinfo{author}{\bibfnamefont{C.}~\bibnamefont{Baines}},
	\bibinfo{journal}{J.Phys.Condens.Matter} \textbf{\bibinfo{volume}{17}},
	\bibinfo{pages}{979} (\bibinfo{year}{2005}).
	
	\bibitem[{\citenamefont{Sarte et~al.}(2011)\citenamefont{Sarte, Silverstein,
			Wyk, Gardner, Qiu, Zhou, and Wiebe}}]{Sarte2011}
	\bibinfo{author}{\bibfnamefont{P.~M.} \bibnamefont{Sarte}},
	\bibinfo{author}{\bibfnamefont{H.~J.} \bibnamefont{Silverstein}},
	\bibinfo{author}{\bibfnamefont{B.~T. K.~V.} \bibnamefont{Wyk}},
	\bibinfo{author}{\bibfnamefont{J.~S.} \bibnamefont{Gardner}},
	\bibinfo{author}{\bibfnamefont{Y.}~\bibnamefont{Qiu}},
	\bibinfo{author}{\bibfnamefont{H.~D.} \bibnamefont{Zhou}}, \bibnamefont{and}
	\bibinfo{author}{\bibfnamefont{C.~R.} \bibnamefont{Wiebe}},
	\bibinfo{journal}{J.Phys.Condens.Matter} \textbf{\bibinfo{volume}{23}},
	\bibinfo{pages}{382201} (\bibinfo{year}{2011}).
	
	\bibitem[{\citenamefont{Mydosh}(1993)}]{Mydosh1993}
	\bibinfo{author}{\bibfnamefont{J.~A.} \bibnamefont{Mydosh}},
	\emph{\bibinfo{title}{Spin Glasses: An Experimental Introduction}}
	(\bibinfo{publisher}{Taylor and Francis, London}, \bibinfo{year}{1993}).
	
	\bibitem[{\citenamefont{Palmer and Chalker}(2000)}]{Palmer2000}
	\bibinfo{author}{\bibfnamefont{S.~E.} \bibnamefont{Palmer}} \bibnamefont{and}
	\bibinfo{author}{\bibfnamefont{J.~T.} \bibnamefont{Chalker}},
	\bibinfo{journal}{Phys.Rev.B} \textbf{\bibinfo{volume}{62}},
	\bibinfo{pages}{488} (\bibinfo{year}{2000}).
	
	\bibitem[{\citenamefont{Hallas et~al.}(2016)\citenamefont{Hallas, Gaudet,
			Butch, Tachibana, Freitas, Luke, Wiebe, and Gaulin}}]{Hallas2016}
	\bibinfo{author}{\bibfnamefont{A.~M.} \bibnamefont{Hallas}},
	\bibinfo{author}{\bibfnamefont{J.}~\bibnamefont{Gaudet}},
	\bibinfo{author}{\bibfnamefont{N.~P.} \bibnamefont{Butch}},
	\bibinfo{author}{\bibfnamefont{M.}~\bibnamefont{Tachibana}},
	\bibinfo{author}{\bibfnamefont{R.~S.} \bibnamefont{Freitas}},
	\bibinfo{author}{\bibfnamefont{G.~M.} \bibnamefont{Luke}},
	\bibinfo{author}{\bibfnamefont{C.~R.} \bibnamefont{Wiebe}}, \bibnamefont{and}
	\bibinfo{author}{\bibfnamefont{B.~D.} \bibnamefont{Gaulin}},
	\bibinfo{journal}{Phys.Rev.B} \textbf{\bibinfo{volume}{93}},
	\bibinfo{pages}{100403(R)} (\bibinfo{year}{2016}).
	
	\bibitem[{\citenamefont{Shirai}(2007)}]{Masae2007}
	\bibinfo{author}{\bibfnamefont{M.}~\bibnamefont{Shirai}}, Ph.D. thesis,
	\bibinfo{school}{University College London} (\bibinfo{year}{2007}).
	
	\bibitem[{\citenamefont{Petit et~al.}(2016)\citenamefont{Petit, Lhotel, Canals,
			Hatnean, Ollivier, 4, Ressouche, Wildes, Lees, and Balakrishnan}}]{Petit2016}
	\bibinfo{author}{\bibfnamefont{S.}~\bibnamefont{Petit}},
	\bibinfo{author}{\bibfnamefont{E.}~\bibnamefont{Lhotel}},
	\bibinfo{author}{\bibfnamefont{B.}~\bibnamefont{Canals}},
	\bibinfo{author}{\bibfnamefont{M.~C.} \bibnamefont{Hatnean}},
	\bibinfo{author}{\bibfnamefont{J.}~\bibnamefont{Ollivier}},
	\bibinfo{author}{\bibfnamefont{H.~M.} \bibnamefont{4}},
	\bibinfo{author}{\bibfnamefont{E.}~\bibnamefont{Ressouche}},
	\bibinfo{author}{\bibfnamefont{A.~R.} \bibnamefont{Wildes}},
	\bibinfo{author}{\bibfnamefont{M.~R.} \bibnamefont{Lees}}, \bibnamefont{and}
	\bibinfo{author}{\bibfnamefont{G.}~\bibnamefont{Balakrishnan}},
	\bibinfo{journal}{Nat. Phys.} \textbf{\bibinfo{volume}{12}},
	\bibinfo{pages}{746} (\bibinfo{year}{2016}).
	
	\bibitem[{\citenamefont{Brooks-Bartlett
			et~al.}(2014)\citenamefont{Brooks-Bartlett, Banks, Jaubert, Harman-Clarke,
			and Holdsworth}}]{BrooksBartlett2014}
	\bibinfo{author}{\bibfnamefont{M.~E.} \bibnamefont{Brooks-Bartlett}},
	\bibinfo{author}{\bibfnamefont{S.~T.} \bibnamefont{Banks}},
	\bibinfo{author}{\bibfnamefont{L.~D.~C.} \bibnamefont{Jaubert}},
	\bibinfo{author}{\bibfnamefont{A.}~\bibnamefont{Harman-Clarke}},
	\bibnamefont{and} \bibinfo{author}{\bibfnamefont{P.~C.~W.}
		\bibnamefont{Holdsworth}}, \bibinfo{journal}{Phys.Rev.X}
	\textbf{\bibinfo{volume}{4}}, \bibinfo{pages}{011007} (\bibinfo{year}{2014}).
	
	\bibitem[{\citenamefont{Yan et~al.}(2017)\citenamefont{Yan, Benton, Jaubert,
			and Shannon}}]{Yan2017}
	\bibinfo{author}{\bibfnamefont{H.}~\bibnamefont{Yan}},
	\bibinfo{author}{\bibfnamefont{O.}~\bibnamefont{Benton}},
	\bibinfo{author}{\bibfnamefont{L.}~\bibnamefont{Jaubert}}, \bibnamefont{and}
	\bibinfo{author}{\bibfnamefont{N.}~\bibnamefont{Shannon}},
	\bibinfo{journal}{Phys.Rev.B} \textbf{\bibinfo{volume}{95}},
	\bibinfo{pages}{094422} (\bibinfo{year}{2017}).
	
	\bibitem[{\citenamefont{Zhitomirsky et~al.}(2014)\citenamefont{Zhitomirsky,
			Holdsworth, and Moessner}}]{Zhitomirsky2014}
	\bibinfo{author}{\bibfnamefont{M.~E.} \bibnamefont{Zhitomirsky}},
	\bibinfo{author}{\bibfnamefont{P.~C.~W.} \bibnamefont{Holdsworth}},
	\bibnamefont{and} \bibinfo{author}{\bibfnamefont{R.}~\bibnamefont{Moessner}},
	\bibinfo{journal}{Phys.Rev.B} \textbf{\bibinfo{volume}{89}},
	\bibinfo{pages}{140403(R)} (\bibinfo{year}{2014}).
	
	\bibitem[{\citenamefont{Vegard}(1921)}]{Vegard1921}
	\bibinfo{author}{\bibfnamefont{L.}~\bibnamefont{Vegard}}, \bibinfo{journal}{Z.
		Physik} \textbf{\bibinfo{volume}{5}}, \bibinfo{pages}{17}
	(\bibinfo{year}{1921}).
	
\end{thebibliography}
\end{document}